\documentstyle[epsf]{mn}

\newif\ifAMStwofonts

\newcommand{\gsim}{\:\raisebox{.25ex}{$>$}\hspace*{-.75em}
      \raisebox{-.93ex}{$\sim$}\:}
\newcommand{\lsim}{\:\raisebox{.25ex}{$<$}\hspace*{-.75em}
      \raisebox{-.93ex}{$\sim$}\:}
\newcommand{\xmn}[2]{\mbox{$#1\!\times\! 10^{#2}\,$}}
\newcommand{\NMS}[1]{\mbox{$#1\,{\cal M}_{\textstyle\odot}$}} 
\newcommand{\isn}[2]{\mbox{$^{#2}${#1}}} 
\ifoldfss
  \ifCUPmtlplainloaded \else
    \NewTextAlphabet{textbfit} {cmbxti10} {}
    \NewTextAlphabet{textbfss} {cmssbx10} {}
    \NewMathAlphabet{mathbfit} {cmbxti10} {} 
    \NewMathAlphabet{mathbfss} {cmssbx10} {} 
  \fi
  \ifAMStwofonts
    \ifCUPmtlplainloaded \else
      \NewSymbolFont{upmath} {eurm10}
      \NewSymbolFont{AMSa} {msam10}
      \NewMathSymbol{\upi}     {0}{upmath}{19}
      \NewMathSymbol{\umu}     {0}{upmath}{16}
      \NewMathSymbol{\upartial}{0}{upmath}{40}
      \NewMathSymbol{\leqslant}{3}{AMSa}{36}
      \NewMathSymbol{\geqslant}{3}{AMSa}{3E}

    \fi
  \fi
\fi 

\ifnfssone
  \newmathalphabet{\mathit}
  \addtoversion{normal}{\mathit}{cmr}{m}{it}
  \addtoversion{bold}{\mathit}{cmr}{bx}{it}
  \newmathalphabet{\mathbfit} 
  \addtoversion{normal}{\mathbfit}{cmr}{bx}{it}
  \addtoversion{bold}{\mathbfit}{cmr}{bx}{it}
  \newmathalphabet{\mathbfss} 
  \addtoversion{normal}{\mathbfss}{cmss}{bx}{n}
  \addtoversion{bold}{\mathbfss}{cmss}{bx}{n}
  \ifAMStwofonts
    \ifCUPmtlplainloaded \else
      \UseAMStwoboldmath
      \makeatletter
      \new@mathgroup\upmath@group
      \define@mathgroup\mv@normal\upmath@group{eur}{m}{n}
      \define@mathgroup\mv@bold\upmath@group{eur}{b}{n}
      \edef\UPM{\hexnumber\upmath@group}
      \new@mathgroup\amsa@group
      \define@mathgroup\mv@normal\amsa@group{msa}{m}{n}
      \define@mathgroup\mv@bold\amsa@group{msa}{m}{n}
      \edef\AMSa{\hexnumber\amsa@group}
      \makeatother
      \mathchardef\upi="0\UPM19
      \mathchardef\umu="0\UPM16
      \mathchardef\upartial="0\UPM40
      \mathchardef\leqslant="3\AMSa36
      \mathchardef\geqslant="3\AMSa3E
    \fi
  \fi
\fi 

\ifnfsstwo
  \DeclareMathAlphabet{\mathbfit}{OT1}{cmr}{bx}{it}
  \SetMathAlphabet\mathbfit{bold}{OT1}{cmr}{bx}{it}
  \DeclareMathAlphabet{\mathbfss}{OT1}{cmss}{bx}{n}
  \SetMathAlphabet\mathbfss{bold}{OT1}{cmss}{bx}{n}
  \ifAMStwofonts
    \ifCUPmtlplainloaded \else
      \DeclareSymbolFont{UPM}{U}{eur}{m}{n}
      \SetSymbolFont{UPM}{bold}{U}{eur}{b}{n}
      \DeclareSymbolFont{AMSa}{U}{msa}{m}{n}
      \DeclareMathSymbol{\upi}{0}{UPM}{"19}
      \DeclareMathSymbol{\umu}{0}{UPM}{"16}
      \DeclareMathSymbol{\upartial}{0}{UPM}{"40}
      \DeclareMathSymbol{\leqslant}{3}{AMSa}{"36}
      \DeclareMathSymbol{\geqslant}{3}{AMSa}{"3E}
    \fi
  \fi
\fi 

\ifCUPmtlplainloaded \else
  \ifAMStwofonts \else 
    \def\upi{\pi}
    \def\umu{\mu}
    \def\upartial{\partial}
  \fi
\fi

\title{Explosion Energies, Nickel Masses, and Distances of
        Supernovae of Type IIP}
\author[D. K. Nadyozhin]
       {D. K. Nadyozhin \\
 Institute of Theoretical and Experimental Physics, Moscow, 117259, Russia\\
 Max-Planck-Institut f\"ur Astrophysik, Garching, 85741, Germany\\
 Astronomisches Institut der Universit\"{a}t Basel, Binningen, CH-4102, Switzerland
     }%
\date{Accepted ????.
      Received 2003 April 28;
      in original form 2003 April 28}

\pagerange{\pageref{firstpage}--\pageref{lastpage}}
\pubyear{2003}

\begin{document}

\maketitle

\label{firstpage}

\begin{abstract}
  The hydrodynamical modelling of Type II plateau supernova
  light curves predicts a correlation between three observable
  parameters (the plateau duration, the absolute magnitude and
  photospheric velocity at the middle of the plateau)
  on the one side and three physical parameters
  (the explosion energy $E$, the mass of the envelope 
  expelled ${\cal M}$, and the presupernova radius $R$) on the 
  other side. The correlation is used, together with {\em adopted\/}
  EPM distances, to estimate $E$, ${\cal M}$,
  and $R$ for a dozen of well-observed SNe$\,$IIP.
  For this set of supernovae, the resulting value of $E$
   varies within a factor of 6 
  $\left(0.5\lsim E/10^{51}\mbox{erg}\lsim 3\right)$, whereas 
  the envelope mass remains within 
  the limits $10\lsim {\cal M}/{\cal M}_{\textstyle\odot}\lsim 30$.
  The presupernova radius is typically $(200-600)\, R_{\textstyle\odot}$,
  but can reach $\gsim 1000\, R_{\textstyle\odot}$
  for the brightest supernovae (e.g., SN$\,$1992am).
  
  A new method of determining the distance of SNe$\,$IIP
  is proposed. It is based on the assumption of a
  correlation between the explosion energy $E$ and the 
  \isn{Ni}{56} mass required to power the post-plateau light 
  curve tail through \isn{Co}{56} decay.
  The method is useful for SNe$\,$IIP with well-observed 
  bolometric light curves both during
  the plateau and radioactive tail phases.
  The resulting distances and future improvements are discussed.
\end{abstract}


\section{Introduction}
 Plateau Type II supernovae (SNe$\,$IIP) are believed to 
 come from the explosion of massive supergiant stars whose 
 envelopes are rich in hydrogen. Their light curves are easy
 to identify by a long plateau (sometimes up to 120--150 d)
 which is the result of the propagation of a cooling-and-recombination
 wave (CRW) through the supernova envelope that is in a state of free
 inertial expansion $(u=r/t)$. The CRW physics is discussed in detail 
 by Imshennik \& Nadyozhin (1964), 
 Grassberg, Imshennik \& Nadyozhin (1971), and 
 Grassberg \& Nadyozhin (1976). The CRW propagates supersonically
 downward through the expanding supernova envelope and separates 
 almost recombined outer layers from still strongly ionized inner ones. 
 During the plateau phase, the photosphere sits on the upper edge 
 of the CRW front. Since the CRW downward speed turns out to be close
 to the velocity of the outward expansion, the photospheric radius changes
 only slowly during the plateau phase. If one takes into account that also
 the effective temperature does not change appreciably (it approximately
 equals the recombination temperature 5000--7000$\,$K), 
 the approximate constancy of the luminosity becomes obvious.
 
  The supernova outburst properties are determined mainly by 
  three physical parameters: the explosion energy $E$, the mass 
  ${\cal M}$ of the envelope expelled, and the initial radius 
  $R$ of the star just before the explosion (presupernova).
  Litvinova \& Nadyozhin (1983, 1985)
  have undertaken an attempt to derive these parameters from
  a comparison of the hydrodynamical supernova models with 
  observations. They constructed simple approximation formulae 
  which allow to estimate $E$, ${\cal M}$, and $R$ from the 
  observations of individual SNe$\,$IIP. Their results were 
  confirmed by an independent semi-analytical study (Popov 1993).
  At that time, only one or two supernovae were
  sufficiently observed to apply these formulae. At present,
  there exist detailed observational data for 14 such supernovae, 
  including in 12 cases expanding photosphere (EPM) distances,
  which we use in section 2 to estimate 
  $E$, ${\cal M}$, and $R$ by means of these formulae.
   
  In section 3, we propose a new method of distance determination 
  and employ it to 9 individual SNe$\,$IIP which are 
  well-observed both at the plateau and radioactive-tail phases. 
  The method is based on the assumption of
  a correlation between the explosion energy $E$ and the mass 
  of \isn{Ni}{56} in the supernova envelope.
  In section 4 we compare physical parameters and distances 
  of SNe$\,$IIP as derived from the new method 
  with those obtained previously from the EPM method
  and discuss also other aspects of our results.
  Concluding remarks are given in section 5. 

  The preliminary results of this study were reported to
  the Workshop on physics of supernovae held at Garching,
  Germany, July 2002 (Nadyozhin 2003).
  
 \section{A comparison of hydrodynamic models with observations}
  
\begin{figure}
   \centering
 \begin{minipage}[t]{0.46\textwidth}
   \epsfxsize=0.98\textwidth
  \epsffile{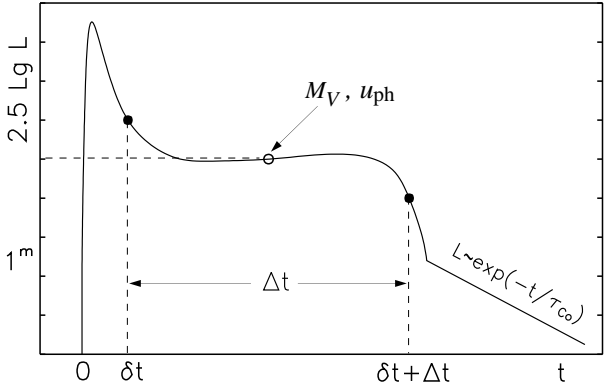}
 \vspace*{-2mm}
 \caption{A schematic SN$\,$IIP light curve. An open circle
  marks the middle of the plateau and two black circles show
  the plateau boundaries. The light curve tail powered by the 
  \protect\isn{Co}{56} decay is also shown 
  $(\tau_{\mathrm{Co}}=111.3\,{\mathrm{d}})$.}
  \label{nadezhfg1}
\end{minipage}
   \end{figure}

  Figure$\,$\ref{nadezhfg1} shows a schematic 
  SNe$\,$IIP light curve. The plateau is defined as part of 
  the light curve on which the supernova brightness remains
  within $1$ mag of the mean value.
  For some supernovae,
  the plateau begins almost immediately after the onset of 
  the explosion $(t=0)$ whereas for others a short luminosity 
  peak can precede the plateau. The peak either appears owing to 
  a shock wave breakout in the case of presupernovae of not very 
  large initial radii $(R\lsim 1000\, R_{\textstyle\odot})$ or,
  according to Grassberg et al. (1971), originates from
  the emergence of a thermal wave precursor for presupernovae 
  of very large radii $(R\approx (2000-5000)\, 
  R_{\textstyle\odot})$ and of moderate explosion energies 
  $(E\lsim\xmn{1}{51}{\mathrm{erg}})$, 
  or at last it may occur as a result of
  interaction between the supernova envelope and a dense stellar 
  wind (Grassberg \& Nadyozhin 1987).
  For some SNe$\,$IIP the peak
  duration $\delta t$ lasts only a few days and is difficult to
  observe (shock wave breakout), for others 
  it could be as large as 10--20 days (thermal wave or dense wind) --
  examples for the latter may be such supernovae as 
  SNe$\,$1988A, 1991al, and 1992af (see below).

 It is quite clear that the middle of the plateau is to be used as
 the main reference point to compare the theoretical models with
 observations. 
 Litvinova \& Nadyozhin (1983, 1985) 
 (LN83 and LN85, hereafter) calculated a grid 
 of the supernova models for $E$, ${\cal M}$, and $R$ within limits
 of  \xmn{(0.18-2.91)}{51}erg, \NMS{(1-16)}, 
 and $(300-5000)\, R_{\textstyle\odot}$. They found 
 $E$, ${\cal M}$, $R$ to be strongly correlated with the plateau
 duration $\Delta t$, and the middle-plateau values of the absolute
 $V$-magnitude $M_V$, and the expansion velocity $u_{\mathrm{ph}}$ 
 at the level of the photosphere (Fig. \ref{nadezhfg1}). According
 to LN85, the following approximate relations can be used to derive 
 $E$, ${\cal M}$, and $R$ from observations:  
\begin{eqnarray}
 \lg E & \!\! =\!\! &\phantom{-}0.135\, M_V + 2.34\lg\Delta t 
 + 3.13\lg u_{\mathrm{ph}} - 4.205\, ,\label{Evtu}\\
 \lg {\cal M} &\!\! =\!\! &\phantom{-}0.234\, M_V + 2.91\lg\Delta t 
 + 1.96\lg u_{\mathrm{ph}} - 1.829\, ,\label{Mvtu}\\
 \lg R & \!\! =\!\! & - 0.572\, M_V - 1.07\lg\Delta t 
 - 2.74\lg u_{\mathrm{ph}} - 3.350\, ,\label{Rvtu}
\end{eqnarray}
where $E$ is expressed in units of $10^{51}\,$erg, ${\cal M}$ and $R$ 
are in solar units, $\Delta t$ in days, and $u_{\mathrm{ph}}$ 
in $1000\,{\mathrm{km}}\,{\mathrm{s}}^{-1}$.
Here $M_V$ can be expressed through the apparent $V$
magnitude by the relation:
 \begin{equation}
  M_V=V-A_V-5\lg(D/1{\mathrm{Mpc}})-25\, , \label{MVAD} 
 \end{equation}
 where $D$ is the distance to a supernova and $A_V$
 is the total absorption on the way to the supernova. One can find
 from Eqs.$\,$(\ref{Evtu})--(\ref{Rvtu}) that $E$, ${\cal M}$, and $R$ 
 scale with the distance as:
 \begin{equation}\label{EMRdist}
  E\sim D^{-0.675},\quad {\cal M}\sim D^{-1.17},\quad R\sim D^{2.86}\, .  
\end{equation}
 Thus, it is very important to know $D$ with as high accuracy as 
 possible. 
 We have selected 14 SNe whose observational data
 are collected in Table$\, 1$. The entries are: the heliocentric
 recession velocities $v_0$ (from the NED: NASA/IPAC Extragalactic Database)
 in column 3, the total absorption
 $A_V$ in column 4, the apparent $V$ magnitude of the mid-point 
 of the plateau in column 5, the duration $\Delta t$ of the plateau
 in column 6, and the photosphere expansion velocity $u_{\mathrm{ph}}$
 in column 7. The references are in column 8.
 
  In order to check the extrapolative capability of
  Eqs.$\,$(\ref{Evtu})--(\ref{Rvtu}), 
  we have included the SN$\,$1987A in our analysis.
  It is well known, that the SN$\,$1987A presupernova radius
  was as small as $\approx 50\, R_{\textstyle\odot}$ -- i.e. outside
  the interval of $(300-5000)\, R_{\textstyle\odot}$ encompassed by
  the above equations. 
  Moreover, the major part of the SN$\,$1987A 
  plateau (about 70 of 110$\,$d) was powered by the \isn{Co}{56}-decay
  (see the review of Imshennik \& Nadyozhin 1989 and 
   references therein).
\begin{table*}
  \centering
 \begin{minipage}{140mm}
  \caption{Observational data for 14 SNe$\,$IIP.}\label{Input}
  \vspace*{2mm}
  \begin{tabular}{llrllrll}\hline
    &  &  &  &  &  &  &  \\[-3mm] 
 \multicolumn{1}{c}{SN} & \multicolumn{1}{c}{Host galaxy}&
 \multicolumn{1}{c}{$v_0$} &
 \multicolumn{1}{c}{$A_V$} & \multicolumn{1}{c}{$V$}& 
 \multicolumn{1}{c}{$\Delta t$} & 
 \multicolumn{1}{c}{$u_{\mathrm{ph}}$} & 
 \multicolumn{1}{c}{Ref.}\\
 &  & \multicolumn{1}{c}{\footnotesize km$\,$s$^{-1}$} &
 \multicolumn{1}{c}{\footnotesize mag} & 
 \multicolumn{1}{c}{\footnotesize mag} & 
 \multicolumn{1}{c}{\footnotesize days} & 
 \multicolumn{1}{c}{\footnotesize km$\,$s$^{-1}$} &  \\
 \multicolumn{1}{c}{\footnotesize (1)} &
 \multicolumn{1}{c}{\footnotesize (2)} &
 \multicolumn{1}{c}{\footnotesize (3)} &
 \multicolumn{1}{c}{\footnotesize (4)} &
 \multicolumn{1}{c}{\footnotesize (5)} &
 \multicolumn{1}{c}{\footnotesize (6)} &
 \multicolumn{1}{c}{\footnotesize (7)} &
 \multicolumn{1}{c}{\footnotesize (8)} \\
    &  &  &  &  &  &  &  \\[-4mm]\hline
    &  &  &  &  &  &  &  \\[-3mm]
    1968L &  NGC 5236 & 516  & $0.219$ &
       $12.0$  & 80  & 4100 &{\footnotesize 1, 2, 3}\\
    1969L &  NGC 1058 & 518  & $0.203$ &
       $13.4$  & 100 & 4000 &{\footnotesize 1, 2}\\
    1986L &  NGC 1559 & 1292 & $0.099$ &
       $14.7$  & 110 & 4000 &{\footnotesize 4}\\
    1988A &  NGC 4579 & 1519 & $0.136$ &
       $15.0$  & 110 & 3000 &{\footnotesize 1, 2, 4, 5, 6}\\
    1989L &  NGC 7339 & 1313 & $1.00$ &
       $16.5$  & 140 & 3000 &{\footnotesize 7, 19}\\
    1990E &  NGC 1035 & 1241 & $1.083$ &
       $16.0$  & 120 & 4000 &{\footnotesize 2, 4, 8, 9}\\
    1991al & LEDA 140858   & 4572 & $0.318$ &
       $17.0$  & 90 & 6000  &{\footnotesize 4}\\
    1992af & ESO 340-G038  & 6000 & $0.171$ &
       $17.3$  & 90 & 6000 &{\footnotesize 4, 7}\\
    1992am & anon 0122-04  & 14600& $0.464$ &
       $19.0$  & 110  & 4800 &{\footnotesize 4, 10}\\
    1992ba & NGC 2082      & 1104 & $0.193$ &
       $15.43$ & 100  & 2900 &{\footnotesize 4, 7}\\
    1999cr & ESO 576-G034  & 6069 & $0.324$ &
       $18.6$  & 100  & 3600 &{\footnotesize 4}\\
    1999em & NGC 1637      & 717  & $0.314$ &
       $14.0$  & 110  & 3000 &{\footnotesize 4, 11, 12, 13, 17}\\
    1999gi & NGC 3184      & 592  & $0.65$   &
       $15.0$  & 110  & 2900 &{\footnotesize 14, 15, 16, 18}\\ 
    1987A  & LMC          & 278  & $0.465$ &
       \phantom{1}$3.3$  & 110 & 2900 &{\footnotesize 4}\\[-4mm]
          &              & &  &  &  &  &  \\ \hline
          &              & &  &  &  &  &  \\[-3mm]          
 \multicolumn{8}{l}{\footnotesize {\sc References}. -- 
    (1) Patat et al. 1993; 
    (2) Schmidt et al. 1992;
    (3) Wood \& Andrews 1974;}\\ 
 \multicolumn{8}{l}{\footnotesize (4) Hamuy 2001; 
    (5) Ruiz-Lapuente et al. 1990; 
    (6) Turatto et al. 1993;
    (7) Schmidt et al.}\\
 \multicolumn{8}{l}{\footnotesize 1994a; 
    (8) Schmidt et al. 1993; 
    (9) Benetti et al. 1994; 
   (10) Schmidt et al. 1994b;}\\ 
 \multicolumn{8}{l}{\footnotesize 
   (11) Hamuy et al. 2001;
   (12) Haynes et al. 1998;
   (13) Baron et al. 2000;
   (14) Schlegel 2001;}\\
 \multicolumn{8}{l}{\footnotesize 
   (15) Smartt et al. 2001;
   (16) Li et al. 2002; 
   (17) Elmhamdi et al. 2003;}\\
 \multicolumn{8}{l}{\footnotesize 
   (18) Leonard et al. 2002b; (19) Pennypacker \& Perlmutter 1989}\\[-4mm]
             &              & &  &  &  &  &  \\ \hline
  \end{tabular}
\end{minipage}
\end{table*}

 Derived properties of the 14 SNe$\,$IIP are in Table 2.
 Column 2 is the recession velocity $v_{220}$ of the supernova 
 (column 1) corrected for a self-consistent Virgocentric infall model
 with a local infall vector of $220\,{\mathrm{km}}\,{\mathrm{s}}^{-1}$
 as described by Kraan-Korteweg (1986). Column 3 gives the
 distance $D_H=v_{220}/H_0$ assuming arbitrarily a value of
 $H_0=60\,{\mathrm{km}}\,{\mathrm{s}^{-1}}\,{\mathrm{Mpc}}^{-1}$.
 For comparison, column 4 gives the distance $D_{\mathrm{EPM}}$ obtained 
 with the use of the expanding photosphere method in the references
 listed at the bottom of the table. The SNe 1991al and 1992af are the
 exception. Owing to the incompleteness of the observational data,
 it is hard to determine the EPM distance to the SN$\,$1991al 
 (Hamuy 2001).
 For the same reason, the EPM distance of 55 Mpc for the SN$\,$1992af
 obtained by Schmidt et al. (1994a) seems to be quite uncertain 
 as pointed out by Hamuy (2001). 
 For these two SNe, we present in column 4 the distances
 calculated by Hamuy (2001) from the CMB 
 redshifts and the Hubble constant 
 $H_0=65\,{\mathrm{km}}\,{\mathrm{s}}^{-1}\,{\mathrm{Mpc}}^{-1}$.
 Columns 5--8 are the absolute magnitude $M_V$ of the mid-point 
 of the plateau, the explosion energy $E$, the mass expelled $\cal M$,
 and the presupernova radius $R$ -- all derived from 
 Eqs.$\,$(\ref{MVAD}), (\ref{Evtu})--(\ref{Rvtu}) for the $D_H$ 
 distances listed in column 3.
 Column 9 gives the mass of \isn{Ni}{56}, 
  ejected by some supernovae, which was estimated by reducing 
  the radioactive-tail luminosities, measured by Hamuy (2001),
  to the distances $D_H$ given in column 3.
\begin{table*}
  \centering
 \begin{minipage}{140mm}
  \caption{The supernova physical properties.}\label{ResM}
  \vspace*{2mm}
  \begin{tabular}{lrrrrcrrl}\hline
    &  &  &  &  &  &  &  & \\[-3mm] 
 \multicolumn{1}{c}{SN}       & 
 \multicolumn{1}{c}{$v_{220}$}&
 \multicolumn{1}{c}{$D_H$}    &
 \multicolumn{1}{c}{$D_{\mathrm{EPM}}$} &
 \multicolumn{1}{c}{$M_V$}    & 
 \multicolumn{1}{c}{$E$}      & 
 \multicolumn{1}{c}{${\cal M}$}      & 
 \multicolumn{1}{c}{$R$}      & 
 \multicolumn{1}{c}{${\cal M}_{\mathrm{Ni0}}$}\\
                                        & 
 \multicolumn{1}{c}{\footnotesize km$\,$s$^{-1}$} &
 \multicolumn{1}{c}{\footnotesize Mpc} &
 \multicolumn{1}{c}{\footnotesize Mpc} &
 \multicolumn{1}{c}{\footnotesize mag} &
 \multicolumn{1}{c}{\footnotesize $10^{51}$erg} & 
 \multicolumn{1}{c}{\footnotesize ${\cal M}_{\textstyle\odot}$} &
 \multicolumn{1}{c}{\footnotesize $R_{\textstyle\odot}$} &
 \multicolumn{1}{c}{\footnotesize ${\cal M}_{\textstyle\odot}$} \\
 \multicolumn{1}{c}{\footnotesize (1)} &
 \multicolumn{1}{c}{\footnotesize (2)} &
 \multicolumn{1}{c}{\footnotesize (3)} &
 \multicolumn{1}{c}{\footnotesize (4)} &
 \multicolumn{1}{c}{\footnotesize (5)} &
 \multicolumn{1}{c}{\footnotesize (6)} &
 \multicolumn{1}{c}{\footnotesize (7)} &
 \multicolumn{1}{c}{\footnotesize (8)} &
 \multicolumn{1}{c}{\footnotesize (9)} \\
    &  &  &  &  &  &  &  & \\[-4mm]\hline
    &  &  &  &  &  &  &  & \\[-3mm]
    1968L & 291 & 4.85 & 4.5$^{(1)}$ & $-16.65$ &
       0.83  & 10.3  & 286 &  \\
    1969L & 766& 12.77 & 10.6$^{(1)}$ & $-17.33$ &
       1.05  & 13.0 & 595 &  \\
    1986L & 1121 & 18.68  & 16.0$^{(1)}$ & $-16.76$ &
       1.56  & 23.5 & 251 & 0.026 \\
    1988A & 1179 & 19.65  & 20.0$^{(1)}$ & $-16.60$ &
       0.67  & 14.5 & 452 & 0.082 \\
    1989L &  1556 & 25.93 & 17.0$^{(1)}$ & $-16.57$ &
       1.18  & 29.8 & 334 &  \\
    1990E &  1238 & 20.63 & 18.0$^{(1)}$ & $-16.66$ &
       1.98  & 31.9 & 200 & 0.052 \\
    1991al & 4476   & 74.60 & 70.0$^{(2a)}$ & $-17.68$ &
       2.61 & 17.6 & 347 & 0.12 \\
    1992af & 6000  & 100.00 & 83.70$^{(2a)}$ & $-17.87$ &
       2.46 & 15.9 & 445 & 0.24 \\
    1992am & 14600  & 243.33& 180.0$^{(1)}$ & $-18.40$ &
       1.66 & 13.9  & 1321 & 0.36 \\
    1992ba & 1096   & 18.27 & 22.0$^{(2)}$ & $-16.07$ &
       0.57 & 13.7  & 272 & 0.029\\
    1999cr & 6069  & 101.15 & 86.0$^{(2)}$ & $-16.75$ &
       0.90  & 14.5  & 368 & 0.085 \\
    1999em & 743      & 12.38  & 8.20$^{(3)}$ & $-16.78$ &
       0.63 & 13.2  & 569 & 0.058 \\
    1999gi & 707      & 11.78  & 11.10$^{(4)}$ & $-16.01$ &
       0.72 & 18.7  & 226 & 0.025$^{\textstyle\ast}$\\
    1987A & \multicolumn{1}{c}{---}& 0.05
 & 0.05\phantom{$^{(4)}$} & $-15.66$ & 0.80 & 22.6 & 143 & 0.065 \\[-4mm]
 &  &  &  &  &  &  &  &  \\ \hline
 \multicolumn{9}{l}{\footnotesize $^{(1)}$Eastman, Schmidt, \& Kirshner
 1996; $^{(2)}$Hamuy 2001; $^{(2a)}$Hamuy 2001,
 based}\\
 \multicolumn{9}{l}{\footnotesize \phantom{$^{(1)}$}on an adopted value
  of $H_0=65$, see text; $^{(3)}$Leonard et al. 2002a;}\\
 \multicolumn{9}{l}{\footnotesize
  $^{(4)}$Leonard et al. 2002b}\\
   \multicolumn{9}{l}{$^{\textstyle\ast}$\footnotesize Derived
    from $V=17.86$ mag at $t=174.3\,$d 
    (Leonard et al. 2002b), and Hamuy's}\\
   \multicolumn{9}{l}{\phantom{$^{\textstyle\ast)}$}\footnotesize 
     recipe (section 5.3 of his Thesis) to convert $V$ into 
    luminosity $L$.}\\[-4mm]
           &  &  &  &  &  &  &  &  \\ \hline
  \end{tabular}
\end{minipage}
\end{table*}

  For SN$\,$1987A, the resulting values 
  of $E$ and ${\cal M}$ (Table$\,$2) differ no more than by a factor 
  of 1.5 from current estimates based on a detailed study.
  However, the presupernova radius turned out to be 
  too large. This happened because the LN85 approximations do not
  take into account the radioactive heating.  
   An advanced study (Grassberg \& Nadyozhin 1986) demonstrates 
   that the radioactive heating  influences only weakly on $E$ and 
   ${\cal M}$ furnished by Eqs.$\,$(\ref{Evtu})--(\ref{Rvtu}), 
   whereas the $R$-values can be overestimated by a factor of 3.
   In this connection,
   one should have in mind  that for some supernovae the $R$-values 
   from Table$\,$2 can be larger than actual presupernova radii.
     
     According to Table$\,$2, 
   the resulting values of $E$, ${\cal M}$, and $R$
   seem to be reasonable enough: the expelled mass, explosion energy, 
   and presupernova radius remain approximately in limits 
   \NMS{(10-30)}, \xmn{(0.6-2.6)}{51}$\,$erg, and 
   $(200-1300)\, R_{\textstyle\odot}$, respectively. 
   Hamuy (2001) assumed that SNe 1991al, 1992af were 
   discovered several weeks after the explosion. Their plateaus, 
   therefore, could have lasted for $\Delta t\approx 110\,$d. 
   It is quite probable, however, that their peak duration
   was $\delta t\approx 20\,$d 
   for the reasons mentioned above.
   Having this in mind, we have chosen in Table$\,$1
   $\Delta t=90\,$d which results in 
   $u_{\mathrm{ph}}=6000\,$km$\,$s$^{-1}$.
   In the case of $\Delta t =110\,$d we would have to assume
   $u_{\mathrm{ph}}=7000\,$km$\,$s$^{-1}$ and would obtain 
   very large values of $E$ and ${\cal M}$ for both supernovae:
   $E\approx\xmn{7}{51}\,$erg and ${\cal M}\approx\NMS{40}$.
   No other special adjustments of the observational data 
   given in Table$\,$1 were made. 

 \section{Plateau-Tail Distance Determination}
 The SN$\,$IIP light curve tails are believed to be powered by the
 \isn{Co}{56} decay. The temporal behavior of the bolometric 
 luminosity is given by (see, e.g. Nadyozhin 1994):
 \begin{equation}\label{LCo}
  L = \xmn{1.45}{43}\:
  \exp\left(-\frac{t}{\tau_{\mathrm{Co}}}\right)
  {{\cal M}_{\mathrm{Ni0}}\over {\cal M}_{\textstyle\odot}}\:
  \mbox{erg s}^{-1}\, ,
 \end{equation}
 where $t$ is measured from the moment of explosion $(t=0)$,
 ${\cal M}_{\mathrm{Ni0}}$ is the total mass of \isn{Ni}{56} at $t=0$
 which decays with a half-life of 6.10$\,$d into \isn{Co}{56}, 
 and $\tau_{\mathrm{Co}}=111.3\,{\mathrm{d}}$. 
 
 Equation$\,$(\ref{LCo}) can be written in the form
 \begin{equation}\label{MNi}
  {\cal M}_{\mathrm{Ni0}}=\frac{ D^2}{145}\, Q,\qquad 
  Q\equiv F_{41}(t)\exp\left(\frac{t}{111.3}\right),  
\end{equation}
 where ${\cal M}_{\mathrm{Ni0}}$ is in ${\cal M}_{\textstyle\odot}$, 
 $t$ in days and $D$ in Mpc.
 The quantity $F_{41}(t)$ is the bolometric tail luminosity measured
 at time $t$ in units $10^{41}\,{\mathrm{erg}}\,{\mathrm{s}}^{-1}$ 
 under the assumption that the supernova is at distance $D=1\,$Mpc. 
 Equation$\,$(\ref{MNi}) contains a single observational 
 parameter $Q$ which is independent of $t$ and also
 makes no assumption on $D$ as long as $F_{41}$ is fixed
 by observations.
 
 Thus, it is irrelevant at which $t$ the luminosity
 is actually measured -- one has only to be sure that the supernova
 really entered its tail phase. Columns $8-10$ of Table$\,$3 
 give $t$ and corresponding values of $F_{41}(t)$ and $Q$ derived 
 from Hamuy's (2001) Figures 5.7 and 5.8 except SN$\,$1999gi
 for which the values were calculated from the data 
 of Leonard et al. (2002b).
 
 If the value of ${\cal M}_{\mathrm{Ni0}}$ was known, one could 
 easily find the distance $D$ from Eq.$\,$(\ref{MNi}). 
 So, we have to look for a way to estimate ${\cal M}_{\mathrm{Ni0}}$
 independently. It seems reasonable to assume that the supernova
 explosion energy $E$ should correlate with ${\cal M}_{\mathrm{Ni0}}$
 produced during the explosion. This means that
 \begin{equation}\label{EMcor}
  E=f\left({\cal M}_{\mathrm{Ni0}}\right)=
  f\left(\frac{ D^2}{145}\, Q\right),
 \end{equation}
 where $f$ represents {\em a statistically admissible correlation 
 function\/} rather than a strict mathematical relation. 
 Inserting this expression for $E$ into Eq.$\,$(\ref{Evtu}) and using
 Eq.$\,$(\ref{MVAD}) for $M_V$, we obtain an equation which can be 
 solved for $D$ when $V-A_V$, $u_{\mathrm{ph}}$, 
 $\Delta t$, and $Q$ are known from observations. Then for given $D$,
 we can find $E$, ${\cal M}$, $R$, and ${\cal M}_{\mathrm{Ni0}}$ from 
 Eqs.$\,$(\ref{Evtu})--(\ref{Rvtu}), Eq.$\,$(\ref{MVAD}), and
 Eq.$\,$(\ref{MNi}), respectively.
  
  What can be said about the function 
  $f\left({\cal M}_{\mathrm{Ni0}}\right)$
  at present, when the details of the SN$\,$II mechanism remain 
  still ambiguous? First of all, it is reasonable to assume that 
  a good fraction of $E$ comes from the recombination of free
  neutrons and protons into \isn{Ni}{56} just at the bottom 
  of the envelope to be finally expelled (Nadyozhin 1978, 
  Bethe 1996).  
  The hydrodynamical modelling of the collapse
  (Nadyozhin 1978) have indicated that under favourable 
  conditions a neutron-proton shell could be accumulated 
  just under the steady accreting shock wave.
  When the mass of such a shell reaches some critical value
  (presumably of the order of $\approx$\NMS{0.1}) the shell
  can become unstable in respect to recombining into 
  the ''iron group" elements (specifically 
  into \isn{Ni}{56}) to supply the stalled shock wave 
  with the energy of $\approx 10^{51}$erg 
  necessary to trigger the supernova. 
  Here, there is a physical analogy 
  with the origin of planetary nebulae from red giants where
  the energy from the recombination of hydrogen 
  and helium causes the expulsion of a red giant
  rarefied envelope.
  The recent study (Imshennik 2002, and references therein)
  of the ''neutrino crown" 
  -- the region enclosed within neutrinosphere and accreting shock,
  turns out to be in line with such a picture of the supernova 
  mechanism .
  However, some Ni can be produced through the explosive 
  carbon-oxygen burning induced by the outgoing shock wave.
  In this case the energy release per unit Ni mass
  is lower by an order of magnitude than for the neutron-proton
  recombination.
  
  The energy released by the neutron-proton recombination, 
  producing a \isn{Ni}{56} mass of ${\cal M}_{\mathrm{Ni0}}$, 
  is given by
 \begin{equation}\label{EnpNi}
    E({\mathrm{np}}\rightarrow{\mathrm{Ni}})\, =\, \xmn{1.66}{52}
\frac{{\cal M}_{\mathrm{Ni0}}}{{\cal M}_{\textstyle\odot}}\,
{\mathrm{erg}}\, .
\end{equation}
  Thus, the production of only $\sim\NMS{0.06}$ of \isn{Ni}{56}
  is sufficient
  to provide the standard explosion energy of $10^{51}\,$erg.
  The current hydrodynamic models of the SN$\,$II explosions
  (Woosley \& Weaver 1995; Rauscher et al. 2002) do not 
  show a correlation between $E$ and ${\cal M}_{\mathrm{Ni0}}$
  because in these models \isn{Ni}{56} comes from explosive 
  silicon and carbon-oxygen burning near to the envelope 
  bottom and its yield is sensitive to the mass cut point. 
  The photometrical and 
  spectroscopical properties of the SN models
  are virtually independent of the mass cut.
  On the contrary, the
  nucleosynthesis yields are very sensitive to 
  the mass cut. In the current SN models 
  the explosion is usually simulated by 
  locating a piston at the internal boundary
  $m={\cal M}_{\mathrm {cut}}$.
  The piston moves with time according to
  a prescribed law, $R_{\mathrm {pis}}(t)$, with the velocity 
  $(\dot R)$ amplitude being chosen to ensure the final 
  kinetic energy of the expelled envelope 
  of the order of $10^{51}$erg. There are two major 
  uncertainties at this point. 
  First, for a given velocity amplitude 
  the resulting nuclear yields are still sensitive
  to the form of the function $R_{\mathrm {pis}}(t)$.
  Second, the presupernova structure (especially chemical 
  composition) in the vicinity of $m={\cal M}_{\mathrm {cut}}$ 
  will always remain ambiguous until the detailed
  mechanism of the SN disintegration onto 
  the collapsed core and thrown envelope is 
  established. The point is that such 2D effects
  as rotation and large-scale mixing can result in
  the presupernova structure different from that
  predicted by the spherically symmetrical models.
  Under such circumstances, it is difficult to find 
  a serious argument against the possibility to
  expel a noticeable amount of \isn{Ni}{56} from
  the recombination of the neutron-proton shell.
  Thus, we propose a neutron-proton layer which 
  is located somewhat deeper than the value of 
  ${\cal M}_{\mathrm {cut}}$ assumed in the current SN models.
  This layer recombines into \isn{Ni}{56} providing the energy 
  sufficient to convert a steady-state accretion shock into the
  outgoing blast wave. In this case a good correlation between
  $E$ and ${\cal M}_{\mathrm{Ni0}}$ is to be expected.
\begin{table*}
  \centering
 \begin{minipage}{140mm}
  \caption{The tail-calibrated supernova physical properties
   $(\xi = 1)$.}\label{ResD}
  \vspace*{2mm}
  \begin{tabular}{lrrcrrlcll}\hline
  &  &  &  &  &  &  & & &\\[-3mm] 
 \multicolumn{1}{c}{SN}       & 
 \multicolumn{1}{c}{$D_{\mathrm{P-T}}$} &
 \multicolumn{1}{c}{$M_V$}    & 
 \multicolumn{1}{c}{$E$}      & 
 \multicolumn{1}{c}{${\cal M}$}      & 
 \multicolumn{1}{c}{$R$}      & 
 \multicolumn{1}{c}{${\cal M}_{\mathrm{Ni0}}$} &
 \multicolumn{1}{c}{$F_{41}(t)$} & 
 \multicolumn{1}{c}{$t$}&
 \multicolumn{1}{c}{$Q$} \\
                                        & 
 \multicolumn{1}{c}{\footnotesize Mpc}  &
 \multicolumn{1}{c}{\footnotesize mag}  &
 \multicolumn{1}{c}{\footnotesize $10^{51}$erg} & 
 \multicolumn{1}{c}{\footnotesize ${\cal M}_{\textstyle\odot}$} &
 \multicolumn{1}{c}{\footnotesize $R_{\textstyle\odot}$ } &
 \multicolumn{1}{c}{\footnotesize ${\cal M}_{\textstyle\odot}$} &
 \multicolumn{1}{c}{\footnotesize $10^{41}$erg/(s$\,$Mpc$^2$)}$\!\!\!\! $& 
 \multicolumn{1}{c}{\footnotesize days } &\\
 \multicolumn{1}{c}{\footnotesize (1)} &
 \multicolumn{1}{c}{\footnotesize (2)} &
 \multicolumn{1}{c}{\footnotesize (3)} &
 \multicolumn{1}{c}{\footnotesize (4)} &
 \multicolumn{1}{c}{\footnotesize (5)} &
 \multicolumn{1}{c}{\footnotesize (6)} &
 \multicolumn{1}{c}{\footnotesize (7)} &
 \multicolumn{1}{c}{\footnotesize (8)} &
 \multicolumn{1}{c}{\footnotesize (9)} &
 \multicolumn{1}{c}{\footnotesize (10)}\\
  &  &  &  &  &  &  &  &  & \\[-4mm]\hline
  &  &  &  &  &  &  &  &  & \\[-3mm]
    1986L$\!\!\!\! $ & 29.67 & $-17.76$ &
 1.14  & 13.7 & 944 & 0.067$\!\!\!\! $ & \xmn{2.25}{-3}$\!\!\!\! $& 180 & 0.0113\\
    1988A$\!\!\!\! $ & 15.21 & $-16.05$ &
 0.79  & 19.6 & 217 & 0.048$\!\!\!\! $ & \xmn{4.96}{-3}$\!\!\!\! $& 200 & 0.0299\\
    1990E$\!\!\!\! $ &  29.16  & $-17.41$ &
 1.57  & 21.3 & 539 & 0.094$\!\!\!\! $ & \xmn{2.67}{-3}$\!\!\!\! $& 200 & 0.0161\\
    1991al$\!\!\!\! $ & 85.31  & $-17.97$ &
 2.38 & 15.0  & 509 & 0.14$\!\!\!\! $ & \xmn{8.13}{-4}$\!\!\!\! $& 140 & 0.00286\\
    1992af$\!\!\!\! $ & 86.45  & $-17.55$ &
 2.71 & 18.8 & 293 & 0.16$\!\!\!\! $ & \xmn{9.02}{-4}$\!\!\!\! $& 140 & 0.00317\\
    1992ba$\!\!\!\! $ & 19.85  & $-16.25$ &
 0.53 & 12.4  & 346 & 0.032$\!\!\!\! $ & \xmn{1.97}{-3}$\!\!\!\! $& 200 & 0.0119\\
    1999em$\!\!\!\! $ & 11.08  & $-16.54$ &
 0.68 & 15.0  & 414 & 0.041$\!\!\!\! $ & \xmn{1.26}{-2}$\!\!\!\! $& 150 & 0.0485\\
    1999gi$\!\!\!\! $ & 14.53  & $-16.46$ &
 0.63 & 14.5  & 411 & 0.038$\!\!\!\! $ & \xmn{5.41}{-3}$\!\!\!\! $& 174 & 0.0259\\
    1987A$\!\!\!\! $  &  0.045 & $-15.42$ &
 0.87 & 25.6 & 104  & 0.053$\!\!\!\! $ & \xmn{8.16}{2}$\!\!\!\! $& 170 & 3762\\[-4mm]
  &  &  &  &  &  &  &  &  & \\ \hline
  \end{tabular}
\end{minipage}
 \end{table*}
 
 The proposed correlation can have a complex nature.
 It is quite probable that the function $f$ in Eq.$\,$(\ref{EMcor})
 depends also on $\cal M$ since the supernova mechanism is expected
 to be sensitive to the presupernova mass.
 For us only the existence of some correlation is important which in
 combination with Eqs.$\,$(\ref{Evtu})--(\ref{Rvtu}) allows to determine
 the distance independently.
 
 To demonstrate how such a method can work we make the simplest
 assumption that $E$ is proportional to 
 $E({\mathrm{np}}\rightarrow{\mathrm{Ni}})$. Then one can write:
 \begin{equation}\label{Ef}
    E\, =\xi\, E({\mathrm{np}}\rightarrow{\mathrm{Ni}})\, =\, 16.6\,\xi
    {\mathcal M}_{\mathrm{Ni0}}\, =\, 0.1145\,\xi\, D^2 \, Q\, ,
\end{equation}
where, as usual, $E$ is in $10^{51}\,{\mathrm{erg}}$, ${\cal M}_{\mathrm{Ni0}}$
in ${\cal M}_{\textstyle\odot}$ and $D$ in Mpc. 
This equation implies that the function
$f$, introduced in Eq.$\,$(\ref{EMcor}), reads as $f(x)=16.6\,\xi\, x$
where $\xi$ is an adjustable parameter which can be either less
or larger than 1. If there is a noticeable contribution to   
${\cal M}_{Ni0}$ from the explosive carbon-oxygen burning
then $\xi < 1$; if a noticeable contribution to the explosion energy
comes from other source rather than
the neutron-proton recombination then $\xi > 1$.

 Inserting $E$ from Eq.$\,$(\ref{Ef}) and $M_V$ from 
 Eq.$\,$(\ref{MVAD}) into Eq.$\,$(\ref{Evtu}) and solving for $D$,
 we obtain:
 \begin{eqnarray}
  \lg D & = & -0.374\,\lg(\xi\, Q)+0.0504\left(V-A_V\right)
  +0.875\,\lg\Delta t \nonumber\\
        &   & + 1.17\,\lg u_{\mathrm{ph}} -2.482\, ,\label{DMNi}
\end{eqnarray}
 where $D$ is in Mpc, $\Delta t$ in days, and $u_{\mathrm{ph}}$ in
 $1000\,{\mathrm{km}}\,{\mathrm{s}}^{-1}$. We will refer to distances
 derived from Eq.$\,$(\ref{DMNi}) as `plateau-tail distances',
 $D_{\mathrm{P-T}}$, hereafter.
 The results are given 
 in Table$\,$3 for nine supernovae selected from Table$\,$2. 
 We did not include SNe 1992am and 1999cr in our analysis
 because their last available observations may not yet reflect
 the radioactive tail phase. Specifically, there are only two 
 observations of SN$\,$1992am at the post-plateau phase
 of the light curve. Since the observations are separated 
 by a short time interval of 3 days, it is difficult to derive
 the inclination of the bolometric light curve with a required
 accuracy to be shure that SN$\,$1992am is already in
 the radioactive-tail phase. Moreover, one has to remember that
 in addition to the Co-decay the tail luminosity can also
 be contributed by the ejecta-wind interaction (see Chugai 1991
 and references therein). SN$\,$1992am is suspicious
 in this respect because its presupernova radius seems to be
 larger than 1000\,$R_{\textstyle\odot}$ (Table$\,$2).
 Hence, the ${\cal M}_{\mathrm{Ni0}}$-values for these SNe 
 in Table$\,$2  could be actually upper limits.
 
 The different columns of Table$\,$3 give the following quantities:
 (2) the distance $D_{\mathrm{P-T}}$ from Eq.$\,$(\ref{DMNi}) setting $\xi=1$;
 (3) the corresponding absolute $V$-magnitude of the mid-point 
     of the plateau $M_V$;
 (4)--(7) the quantities $E$, $\cal M$, $R$, and ${\cal M}_{\mathrm{Ni0}}$
     as in Table$\,$2, but now using the distance $D_{\mathrm{P-T}}$ 
     as in column (1); the columns 
 (8)--(10) are explained above.
 
 The values of $E$, $\cal M$, $R$, and ${\cal M}_{\mathrm{Ni0}}$ 
 for the $\xi$-values different from 1 can be found using the
 following scaling relations which result from 
 Eqs.$\,$(\ref{EMRdist}),
 (\ref{MNi}), and (\ref{DMNi}):
  \begin{equation}\label{EMRxi}
  E\sim \xi^{0.252},\; {\cal M}\sim \xi^{0.438},
  \; R\sim\xi^{-1.07},\;{\cal M}_{\mathrm{Ni0}}\sim\xi^{-0.748}.  
\end{equation}  
 For a fixed $Q$, 
 the dependence of the distance $D_{\mathrm{P-T}}$,
 defined by Eq.$\,$(\ref{DMNi}), on extinction $A_V$ proves
 to be very weak: an error in $A_V$
 of $\pm 1\,$mag changes $D_{\mathrm{P-T}}$ by only $\pm 12\%$.
 However, if the tail luminosity $F_{41}$ is derived from the
 $V$ measurements (just the case of Hamuy's $F_{41}$-values 
 we use here) then the $\lg\, F_{41}$, and consequently
 $\lg\, Q$, scale as $0.4\, A_V$ and 
 $\lg\, D_{\mathrm{P-T}}$, derived from Eq.$\,$(\ref{DMNi}),
 actually varies with $A_V$ in a standard way, as $-0.2\, A_V$. 
 If the tail luminosity were derived from infra-red measurements
 then the resulting $D_{\mathrm{P-T}}$ distances would be 
 largely independent of extinction. 
 Note also rather weak dependence on $\xi Q$:
 $D_{\mathrm{P-T}}\sim (\xi Q)^{-0.374}$.
 For instance, the decrease in $\xi Q$ by 
 a factor of 2 results in an {\em increase\/} 
 of $D_{\mathrm{P-T}}$ by 30\% only.

     The random errors typically of $\pm 10\% $ for
     the $\delta t$ and $u_{\mathrm {ph}}$ values assumed in Table 1
     result in the uncertainty factor of $\approx 1.2$
     for $D_{\mathrm{P-T}}$  and $\approx 1.5$ for
     ${\cal M}_{\mathrm{Ni0}}(\sim D^2)$ given in Table 3. 
     However, one has to keep in mind two main 
     sources of systematic errors: $(i)$ probable 
     deviation of the theoretical models
     (which Eqs. \ref{Evtu}--\ref{Rvtu} are based on) from real SNe,
     and $(ii)$ the presentation of the $E - {\cal M}_{\mathrm{Ni0}}$ 
     correlation in the form of the straight 
     proportionality (Eq. \ref{Ef}). Both the types of
     systematic errors are difficult to estimate
     at present. Although the SN models calculated
     in LN83 and LN85 rest upon a very simplified 
     presupernova structure, they consistently take
     into account the ionization and recombination
     of hydrogen and helium thereby remaining still
     useful. When a new grid of the SN models, based
     on modern evolutionary presupernova structure,
     is created the systematic error $(i)$ certainly 
     will be reduced. The reduction of the 
     systematic error $(ii)$ requires a more profound
     knowledge of the SN mechanism. Empirically, 
     this problem can be solved by adjusting 
     the factor $\xi$ for each individual SN. 
     It is necessary, however, to collect a much 
     more rich statistics (at least by a factor of 3)
     than that available nowadays (only 9 SNe in Table 3).
  
 \section{Discussion}
\begin{figure} 
   \centering
 \begin{minipage}[t]{0.46\textwidth}
    \epsfxsize=0.98\textwidth
  \epsffile{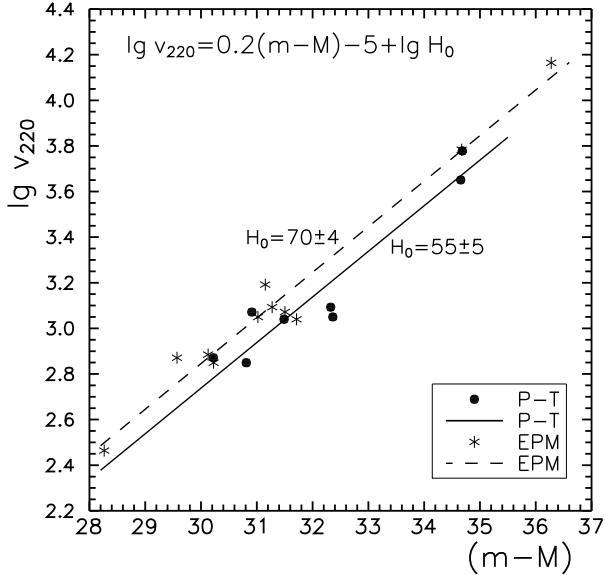}
 \vspace*{-2mm}
 \caption{The Hubble diagram of 8 SNe$\,$IIP with $D_{\mathrm{P-T}}$
  distances from plateau and tail observations (black circles).
  Also shown are the eleven SNe$\,$IIP with known EPM distances
  (asterisks). The respective Hubble lines are fitted to the data.
  The abscissa is the distance modulus $(m-M)=5\lg D+25$.}
  \label{nadezhfg2}
\end{minipage}
   \end{figure}

  The plateau-tail distances derived in section 3 and listed
  in column 2 of Table$\,$3 are plotted in a Hubble diagram in
  Fig.$\,$\ref{nadezhfg2} (except SN$\,$1987A which is not in 
  the Hubble flow). The eight SNe$\,$IIP define a Hubble line with
  $H_0=55\pm 5\,{\mathrm{km}}\,{\mathrm{s}}^{-1}\,{\mathrm{Mpc}}^{-1}$.
  Also shown in Fig.$\,$\ref{nadezhfg2} are the eleven SNe$\,$IIP
  for which EPM distances have been published 
  (column 3 of Table$\,$2). They define a Hubble line of 
  $H_0=70\pm 4\,{\mathrm{km}}\,{\mathrm{s}}^{-1}\,{\mathrm{Mpc}}^{-1}$,
  i.e. the EPM distances are smaller than the plateau-tail distances
  by 25\% on average.
  
   At this point it is not possible to decide which of the two
   results is more nearly correct. Both methods, the plateau-tail
   distances and the EPM distances, depend on assumptions which are
   difficult to verify. The EPM method faces the problem of the
   dilution factor in an expanding atmosphere and the definition
   of the photospheric radius which depends on the uncertainties
   connected with the opacity of an expanding medium.
   However, it may be noted that the EPM distance of SN$\,$1987A 
   agrees well with the generally adopted distance of LMC of 50$\,$kpc
   (Eastman, Schmidt, \& Kirshner  1996)
   and the EPM distance of SN$\,$1968L is indistinguishable
   from the Cepheid distance of NGC$\,$5236 (M83) 
   (Thim et al. 2003).
   
   The main assumption which affects the plateau-tail distances concerns 
   the nature of the proposed $E-{\cal M}_{\mathrm{Ni0}}$ correlation.
   For our simplified example of such a correlation,
   all the uncertainties turn out to be cumulated in the proportionality
   factor $\xi$ between the explosion energy $E$ and the nickel mass
   ${\cal M}_{\mathrm{Ni0}}$. In Table$\,$3 we have adopted a plausible
   value of $\xi=1$, but it cannot be excluded that $\xi$ is 
   as low as 0.5 or as high as 2. Since the Hubble constant scales
   as $H_0\sim\xi^{0.374}$, an average value as high as 
   $\xi=1.9$ would be needed to bring the plateau-tail distances in
   general accord with the EPM distances. Such a high average value of
   $\xi$ is, however, not supported by SNe 1987A and 1999gi.
   If the $D_{\mathrm{P-T}}$ distance of SN$\,$1987A from Table$\,$3
   is scaled to the canonical LMC distance of 50$\,$kpc, $\xi$
   becomes 0.75. And if the host galaxy NGC$\,$3184 of SN$\,$1999gi
   with a $D_{\mathrm{P-T}}$ distance of 14.53$\,$Mpc is a member of
   the same group as NGC$\,$3198 and NGC$\,$3319, for which 
   Freedman et al. (2001) give a mean Cepheid distance of 
   13.5$\,$Mpc, $\xi$ becomes 1.2.
   Eventually additional SNe$\,$IIP with large distances, where 
   the influence of peculiar motions are negligible, will better
   determine the scatter of the Hubble diagram and allow a meaningful
   determination of the actual range of $\xi$.
 
  We have considered three sets of the physical supernova 
  parameters $E$, $\cal M$, and $R$: 
  $(i)$   for the Hubble distances $D_H$
          with $H_0=60\,{\mathrm{km}}\,{\mathrm{s}}^{-1}\,{\mathrm{Mpc}}^{-1}$
          (Table$\,$2, column 3); 
  $(ii)$  for the EPM distances $D_{\mathrm{EPM}}$ (Table$\,$2, column 4); 
  $(iii)$ for the plateau-tail calibrated distances 
          $D_{\mathrm{P-T}}$ (Table$\,$3, column 2).
  \begin{figure} 
 \centering
 \begin{minipage}[t]{0.46\textwidth}
   \epsfxsize=0.95\textwidth
  \epsffile{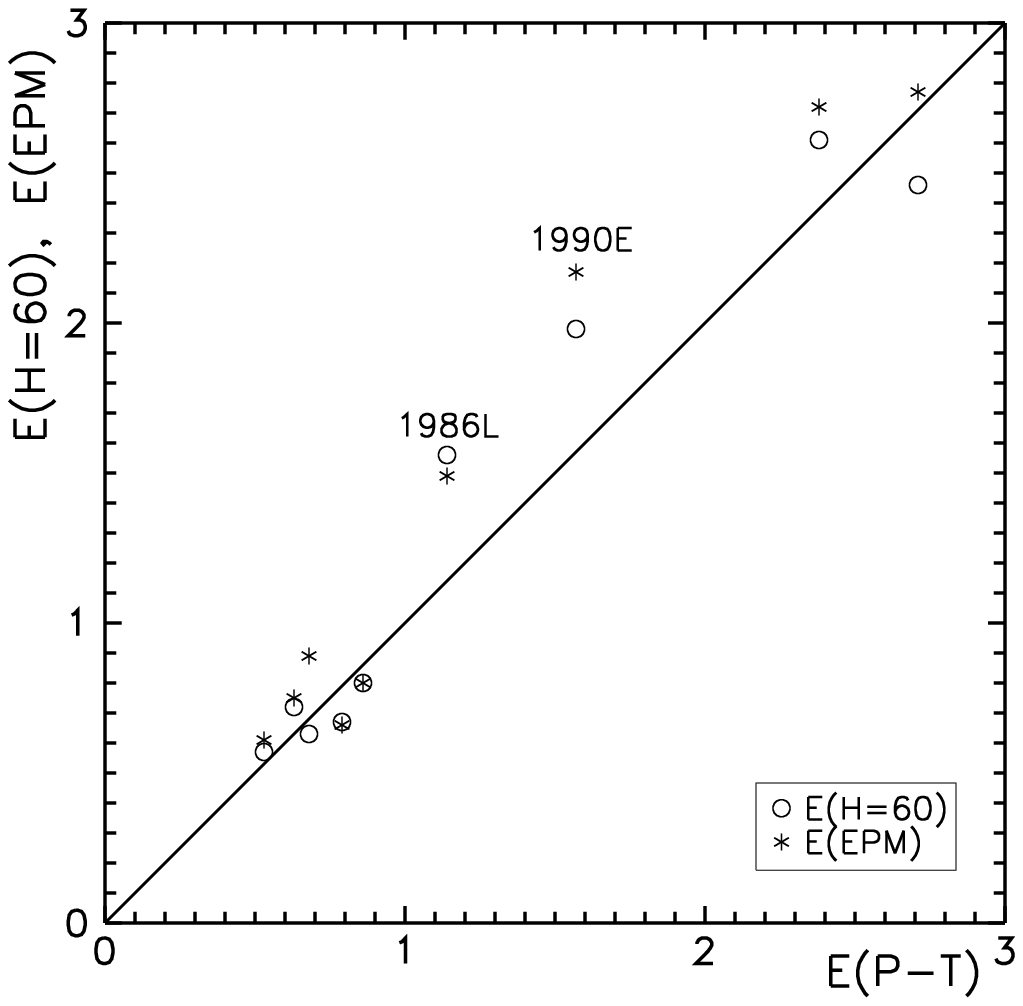}
 \vspace*{-2mm}
 \caption{The explosion energies $E(H_0=60)$ and $E({\mathrm{EPM}})$
          versus $E({\mathrm{P-T}})$ (see the text).}
  \label{nadezhfg3}
 \end{minipage}
 \hspace{\fill}
 \begin{minipage}[t]{0.46\textwidth}
  \epsfxsize=0.95\textwidth
  \epsffile{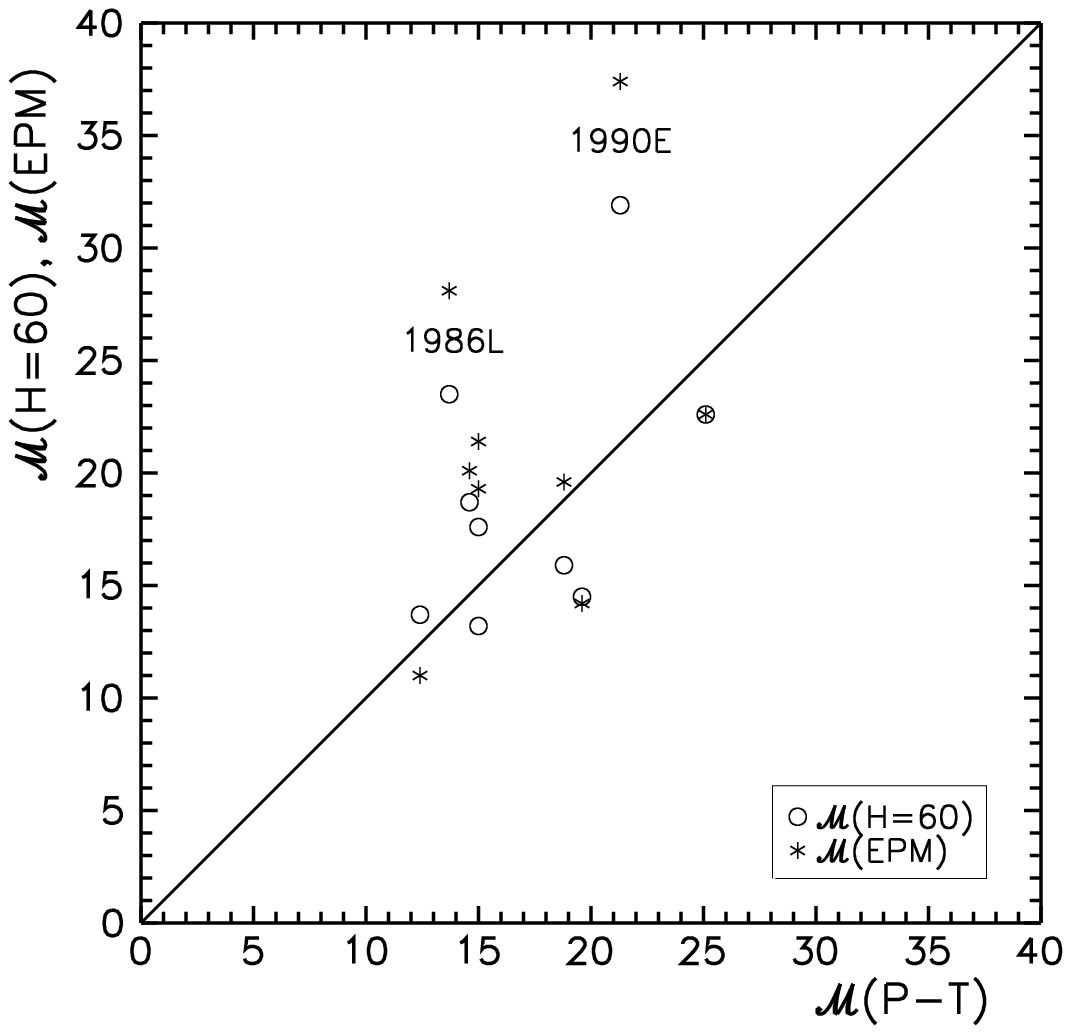}
 \vspace*{-2mm}
 \caption{The expelled masses ${\cal{M}}(H_0=60)$ and 
   $\mathcal{M}({\mathrm{EPM}})$ versus ${\mathcal{M}}({\mathrm{P-T}})$
    (see the text).}
 \label{nadezhfg4}
\end{minipage}
 \end{figure}
 
  Although the above parameters derived from the EPM-distances are
  not presented in Table$\,$2, the corresponding
  $E$ and $\cal M$-values can be read out of 
  Figs.$\,$\ref{nadezhfg3} and \ref{nadezhfg4} which
  compare $E$ and $\cal M$ for sets $(i)$ and $(ii)$ with
  those for set $(iii)$. 
  For seven SNe $E$ and $\cal M$ are rather insensitive 
  to the adopted distances. However for SNe$\,$1986L and 1990E,
  labelled in 
  Figs.$\,$\ref{nadezhfg3} and \ref{nadezhfg4}, the deviations from 
  the (P-T)-values are rather large, especially in case of the envelope
  mass $\cal M$. These SNe differ from others by having a long
  plateau of $(110-120)\,{\mathrm{d}}$ in combination with 
  still a substantial 
  expansion velocity of 4000$\,$km$\,$s$^{-1}$. As a result, 
  their envelope masses $\cal M$, derived from the distances 
  defined by the $D_H$ and 
  $D_{\mathrm{EPM}}$ values, exceed those for other SNe.
  Such a discrepancy for these two SNe 
  is considerably weakened if $\xi\approx 2$.
  Such a high value of $\xi$ implies that half of the explosion energy 
  is supplied
  by a source different from the neutron-proton recombination.
  This may indicate that for massive SNe
  the envelope mass $\cal M$ (in addition to ${\cal M}_{\mathrm{Ni0}}$)
  is involved in the correlation given by Eq.$\,$(\ref{EMcor}).
 
 The random errors of $E$ and $\cal M$ from our approximate 
 Eqs.$\,$(\ref{Evtu})--(\ref{Rvtu}) are estimated to be 
 about $\pm 30$\%$\,$.
 Observational errors
 especially in the expansion velocity $u_{\mathrm{ph}}$
 and the plateau duration $\Delta t$ can
 modify $E$ and ${\cal M}$ by another factor of 1.3. 
 Thus it seems reasonable to assume a random 
 uncertainty of a factor of $\sim 1.5$ for the {\em individual\/} 
 values of $E$ and ${\cal M}$ in Tables$\,$2 and 3.
 The presupernova radii $R$ are very sensitive to distance errors
 (cf. Eq.$\,$\ref{EMRdist}) and may carry random errors of a factor of 2.
 The radii of SNe with large nickel masses like SN$\,$1991al, 1992af
 and perhaps 1992am may carry additional systematic errors 
 because Eqs.$\,$(\ref{Evtu})--(\ref{Rvtu}) 
 do not take into account the radioactive heating in a consistent way. 
   \begin{figure} 
   \centering
 \begin{minipage}[t]{0.46\textwidth}
  \epsfxsize=0.98\textwidth
  \epsffile{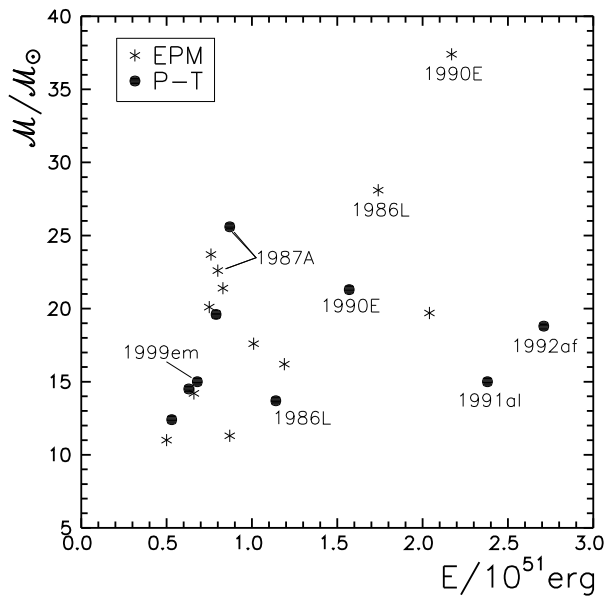}
 \vspace*{-2mm}
 \caption{The explosion energy--envelope mass diagram for 
  the case of EPM distances $D_{\mathrm{EPM}}$ from column 4 of
  Table$\,$2 (asterisks; SNe$\,$1991al and 1992af being excluded)
  and for the case of the plateau-tail distances 
  $D_{\mathrm{P-T}}$ from column 2 of Table$\,$3 (black circles).
  Some SNe are identified (see text).}
  \label{nadezhfg5}
\end{minipage}
   \end{figure}
 
 The expelled masses $\cal M$ are plotted against the explosion
 energies $E$ in Fig.$\,$ \ref{nadezhfg5} for two cases, i.e.
 based on EPM and plateau-tail distances. In case of the 
 $D_{\mathrm{P-T}}$ distances the mean mass of the eight SNe$\,$IIP
 is \NMS{16} with an rms deviation of only \NMS{3}. This narrow mass
 range is contrasted by a wide range of explosion energies
 of \xmn{(0.5-2.7)}{51}erg. The conclusion that there is no
 correlation between the expelled mass -- which is only
 \NMS{(1.4-2)} smaller than the presupernova mass -- and
 the explosion energy is somewhat weakened by the values 
 of $\cal M$ and $E$ based on the EPM distances suggesting
 a marginal correlation between $\cal M$ and $E$ which is
 mainly due to only two SNe: 1986L and 1990E.
 
 One can think of a number of parameters which may explain
 the wide range of explosion energies.
 It could be rotation and magnetic fields inherited by the
 collapsing stellar core. It could be also 
 nonspherical jet-like perturbations of 
 a random nature arising from the macroscopic neutrino-driven 
 advection  below the accretion shock.
 Such perturbations could launch the outgoing
 blast wave earlier when the recombination nuclear energy stored in
 a hot neutron-proton gas was not yet as large as it should be in the
 case of spherical symmetry. If this is correct, one may expect that
 the asphericity of the explosion {\em anticorrelates\/} with the
 explosion energy.
  
 Recently, a promising project has been started 
 (Van Dyk et al. 1999; Smartt et al. 2001, 2002;
  and references therein) with
 the ultimate aim to identify the supernova progenitors (presupernovae)
 or at least to impose conclusive constraints on their masses
 by inspecting the prediscovery field of nearby supernovae. 
 In particular, Smartt et al.  derived upper mass limits
 of \NMS{12} and \NMS{9} for the progenitors of the SNe$\,$1999em
 and 1999gi, assuming distances $D$
 for the host galaxies NGC$\,$1637 and NGC$\,$3184
 of 7.5$\,$Mpc and 7.9$\,$Mpc, respectively. Note that these {\em upper\/}
 limits depend on $D$ and have to be adjusted for other values of $D$ to
 \NMS{12}$\left(D/7.5\,{\mathrm{Mpc}}\right)^{0.6}$ for SN$\,$1999em and 
 \NMS{9}$\left(D/7.9\,{\mathrm{Mpc}}\right)^{0.6}$ for SN$\,$1999gi. 
 This follows
 from the fact that the mass-luminosity relation can be approximated
 as $L\sim {\cal M}^{3.3}$ in the mass interval \NMS{(10-15)}. 
  For SN$\,$1999em at $D_{\mathrm{P-T}}= 11.08\,$Mpc
  (Table$\,$3) follows \NMS{15.2} as the upper mass limit
  for the SN$\,$1999em progenitor. 
  Hence, our result ${\cal M}=\NMS{15.0}$ 
  (Table$\,$3) does not contradict to the observations {\em as long as\/}
  $D({\mathrm{1999em}})\gsim 10\,$Mpc. 
  The situation of SN$\,$1999gi is similar. 
  The upper mass
  limit for $D({\mathrm{1999gi}})=14.53\,$Mpc (Table$\,$3) is
  ${\cal M} < 9\times(14.53/7.9)^{0.6}=\NMS{13.0}$, i.e.  
  not in significant contradiction with the ${\cal M}$-value of 
  \NMS{14.5} from Table$\,$3.
  There is no contradiction either with the upper mass limit of 
  \NMS{15^{+5}_{-3}} for the SN$\,$1999gi progenitor imposed recently 
  by Leonard et al. (2002b). 
  \begin{figure} 
  \centering
 \begin{minipage}[b]{0.46\textwidth}
  \epsfxsize=0.95\textwidth
  \epsffile{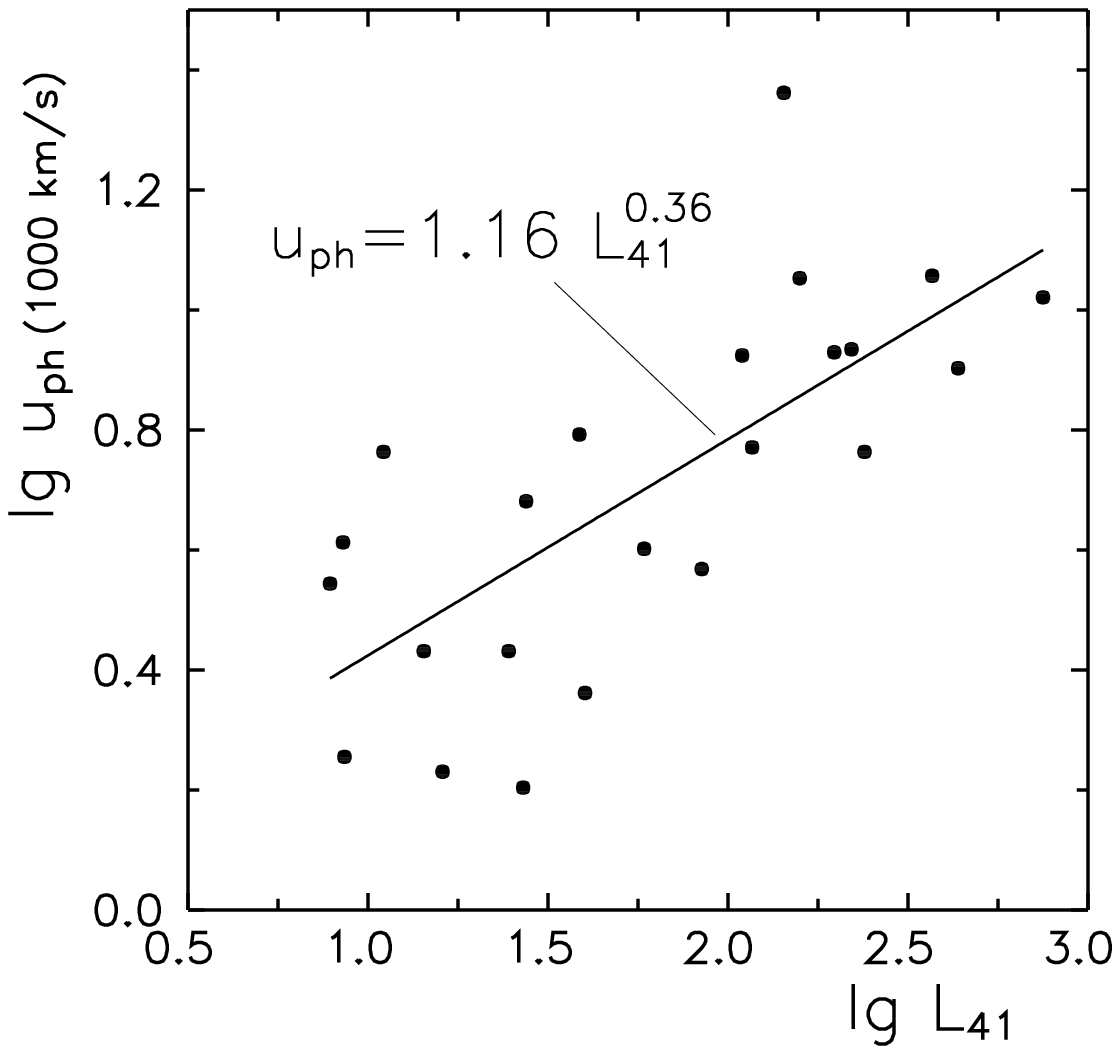}
 \vspace*{-2mm}
 \caption{The correlation of the luminosity $L_{41}$
          (in units $10^{41}\,$erg$\,$s$^{-1}$) of the mid-point
          of the plateau with the expansion velocity $u_{\protect\mathrm{ph}}$
          (in 1000$\,$km$\,$s$^{-1}$) for 23 SN models (black dots).
          }
  \label{nadezhfg6}
\end{minipage}
\hspace{\fill}
\begin{minipage}[b]{0.46\textwidth}
  \epsfxsize=0.95\textwidth
 \vspace*{2mm}
  \epsffile{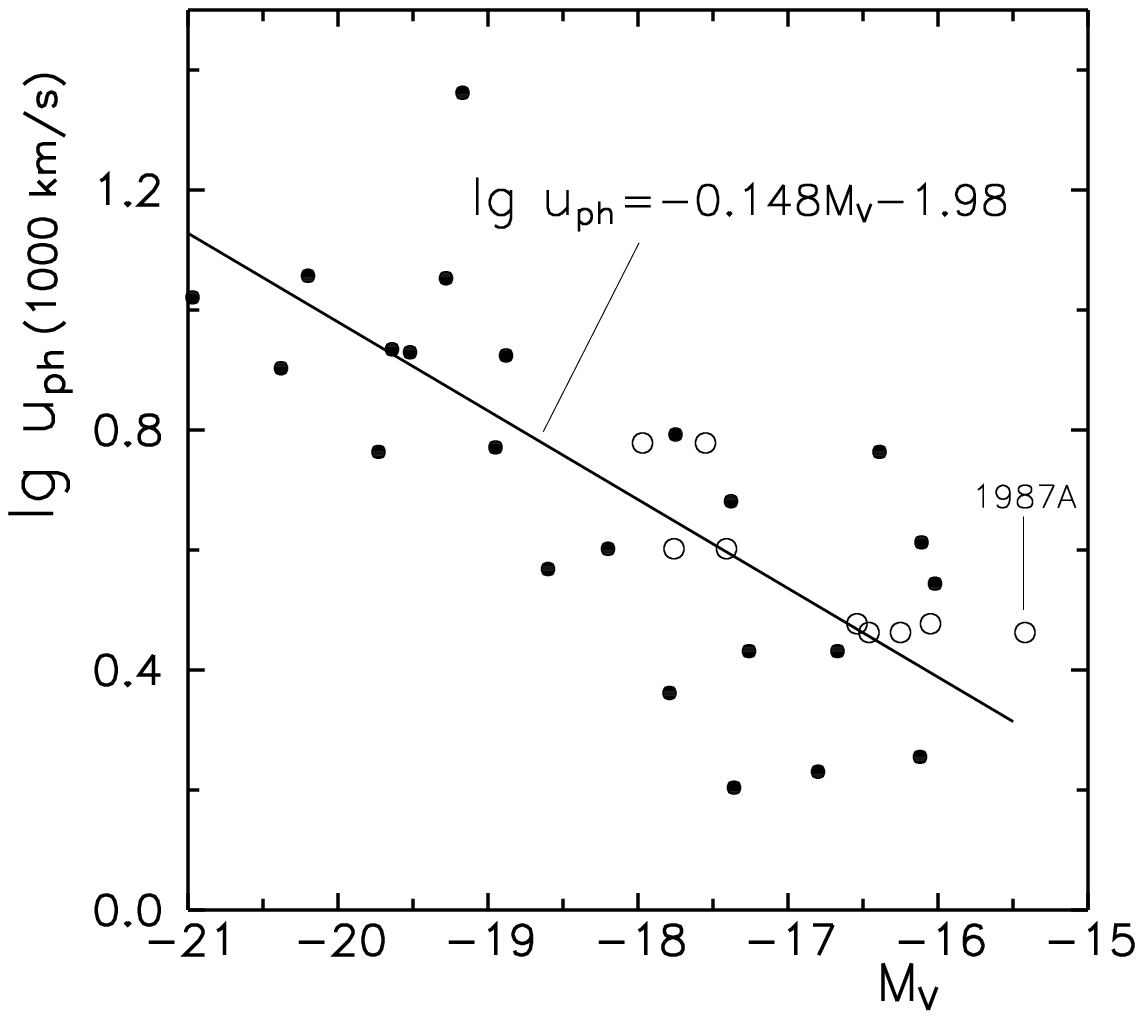}
 \vspace*{-1mm}
 \caption{The correlation of the absolute magnitude $M_V$ of
          the mid-point of the plateau with the expansion velocity
          $u_{\protect\mathrm{ph}}$ for 23 SN models (black dots);
          open circles relate to 9 real SNe including SN$\,$1987A.
         }
 \label{nadezhfg7}
\end{minipage}
\end{figure}

 Equations (\ref{Evtu})--(\ref{Rvtu}) by LN85, derived from
 a grid of 23 SNe$\,$IIP models covering a wide parameter space,
 imply a correlation between the absolute magnitude $M_V$
 (and hence luminosity $L$ -- both measured at the mid-point
 of the plateau) and the expansion velocity $u_{\mathrm{ph}}$.
 The correlation is shown in Figs.$\,$\ref{nadezhfg6} and
 \ref{nadezhfg7} where 23 grid models are shown by black dots;
 the straight lines are the least-squares fits.
 In Fig.$\,$\ref{nadezhfg7} are also shown the eight observed
 SNe$\,$IIP from Table$\,$3 marked by open circles, their
 absolute magnitudes $M_V$ (Table$\,$3, column$\,$2) being
 calculated from Eq.$\,$\ref{MVAD}, where the plateau-tail
 distances $D_{\mathrm{P-T}}$ were used from Table$\,$3, column$\,$2.
 These real SNe follow about the slope of the models, but
 at a fixed value of $u_{\mathrm{ph}}$
 they are fainter by $\approx 0.6\,$mag on average.

 Empirically, Hamuy \& Pinto (2002) have also found,
 using the CMB redshift-based distances, such a correlation.
 The slopes of their least-squares fits are virtually the same as
 shown in Figs.$\,$\ref{nadezhfg6} and \ref{nadezhfg7}.
 Thus our models confirm their finding.

  The main conclusion one can draw from Figs.$\,$\ref{nadezhfg6}
  and \ref{nadezhfg7} is that our three-parametric grid of
  only 23 SNe$\,$IIP properly chosen models is ample enough
  to reproduce the main features of the real SNe.

 \section{Conclusions}
 Model calculation by LN83 and LN85 of SNe$\,$IIP,
 leading to Eqs.$\,$(\ref{Evtu})--(\ref{Rvtu}), are combined
 with available EPM distances and velocity distances $(H_0=60)$
 to derive the explosion energy $E$, the ejected mass $\cal M$,
 and the presupernova radius $R$ of 14 SNe$\,$IIP.
 Only the apparent,
 absorption-corrected magnitude $V$ and the expansion velocity
 $u_{\mathrm{ph}}$ at the mid-point of the plateau together
 with its total duration $\Delta t$ are
 needed as additional input parameters. The results are
 presented in Table$\,$2.

 Instead of using EPM or velocity distances it is also possible
 to use the bolometric fluxes observed during the SN$\,$IIP
 tail phase to determine the Ni mass and hence
 new, independent distances called here
 plateau-tail distances $D_{\mathrm{P-T}}$ (cf. Eq.$\,$\ref{DMNi}).
 The $D_{\mathrm{P-T}}$ distances yield new values of $E$,
 $\cal M$, and $R$ given in Table$\,$3 for nine SNe which were
 observed both during their plateau and tail phases.
 The values of $E$ and $\cal M$, based on EPM and P-T distances
 agree well, with the exception of SNe$\,$1986L and 1990E
 whose masses $\cal M$ coming from P-T distances are by
 a factor of 2 lower than from EPM distances
 (see Fig.$\,$\ref{nadezhfg4}).

  The P-T distances are larger than the EPM distances by
  $\sim 25$\% on average.
  The former suggests a value of $H_0=55\pm 5$.
  The main uncertainty of this result comes from
  the assumption that $\xi=1$, where $\xi$ is the ratio between
  the total explosion energy and the energy liberated by
  the neutron-proton recombination into \isn{Ni}{56}
  (cf. Eq.$\,$\ref{Ef}). To reduce the P-T distances to
  the level of the EPM distances, which correspond to $H_0=70$,
  an average value of $\xi=1.9$ is required. The consequence
  that about half of the total energy $E$ comes from other
  sources than the neutron-proton recombination into \isn{Ni}{56}
  seems rather extreme. In fact it is not supported by
  two SNe$\,$IIP (1987A and 1999gi) with independent distance
  information, which suggest that $\xi$ is of order of unity.
  Moreover, very recently Leonard et al. (2003) have obtained 
   a Cepheid distance of $11.7\pm 1\;$Mpc to NGC 1637 --
   the host galaxy of SN 1999em, which is by a factor
   of 1.4 larger than the EPM distance (Table 2).
   Our $D_{\mathrm{P-T}}$ distance of 11.1 Mpc to SN 1999em 
   (Table 3) is in a good agreement with this result. 
   If it happens
   that the same factor is applicable also to the EPM 
   distances to SNe 1986L and 1990E, there will be 
   no need to resort to large $\xi$-values, such 
   as $\xi\approx 2$ (section 4), to remove the 
   discrepancy between $D_{\mathrm{P-T}}$ and 
   $D_{\mathrm{EPM}}$ for these SNe.

 In conclusion we emphasize the necessity of constructing
 a new grid of hydrodynamic SN$\,$IIP models based on
 current evolutionary presupernova models and taking into
 account \isn{Ni}{56} as an additional parameter in a
 consistent way. Such a grid would allow to create
 more precise analytic
 approximations for a number of correlations between the
 physical parameters of SN$\,$IIP and their observable
 properties.

  The `plateau-tail' method of distance determination
  needs, of course, further critical analysis requiring
  a close collaboration between astronomers observing
  supernovae and theorists modelling their explosions.
  If the proposed $E$--${\cal M}_{\mathrm {Ni}}$
  correlation is confirmed it promises to become a tool 
  to explore the mechanism of SN II with the aid of
  optical and spectroscopical observations.


\section*{Acknowledgments}
 It is a pleasure to express
 my deep gratitude to the Max-Planck-Institut f\"ur Astrophysik
 for hospitality and financial support.
 The work was supported also by the Swiss Science National
 Foundation (Grant 2000/061822) and
 the Russian Foundation for Basic Research (Project 00-02-17230).

 \noindent I am grateful to G.A. Tammann
 for fruitful discussions and a big help and to
 W.~Hillebrandt for useful suggestions.
 I thank B.~Parodi for acquainting me with
 the FORTRAN code converting $v_0$ into $v_{220}$ and
 M.~Hamuy for sending a copy of his Thesis
 where a good portion of the observational data used in this
 work was taken from. The anonymous referee is gratefully
 acknowledged for the constructive critical comments which
 helped to improve the paper.

\bsp

\label{lastpage}

\end{document}